\documentclass[10pt,twocolumn]{IEEEtran}
\usepackage{graphicx}
\usepackage{epsfig}
\usepackage{amsmath}
\usepackage{bbm}
\usepackage{amssymb}
\usepackage{setspace}
\usepackage{wrapfig}
\usepackage{times,color}
\usepackage{caption}
\usepackage{subcaption}
\usepackage{algorithm}
\usepackage{algpseudocode}
\usepackage{epstopdf}
\usepackage{cite,footnote,xspace,syntonly,bm}

\usepackage{amsfonts}

\input{mysymbol.sty}

\algnewcommand{\Inputs}[1]{%
  \State \textbf{Inputs:}
  \Statex \hspace*{\algorithmicindent}\parbox[t]{.8\linewidth}{\raggedright #1}
}
\algnewcommand{\Initialize}[1]{%
  \State \textbf{Initialize:}
  \Statex \hspace*{\algorithmicindent}\parbox[t]{.8\linewidth}{\raggedright #1}
}

 \long\def\symbolfootnote[#1]#2{\begingroup
 	\def\thefootnote{\fnsymbol{footnote}}
 	\footnote[#1]{#2}\endgroup} \psfull

\begin{document}
\title{Nonlinear Structural Vector Autoregressive Models  for  Inferring Effective Brain Network Connectivity}

\author{{{ Yanning Shen, \textit{Student Member}, \textit{IEEE}, Brian Baingana, \textit{Member}, \textit{IEEE},\\
 and Georgios~B.~Giannakis, \textit{Fellow, IEEE}}}\\
}

\markboth{IEEE TRANSACTIONS ON Medical Imaging, \today \; (SUBMITTED)}{}
\maketitle  

\symbolfootnote[0]{$\dag$ Work in this paper was supported by grants NSF 1500713 and NIH 1R01GM104975-01. }

\symbolfootnote[0]{$\ast$ Y. Shen, B. Baingana, and G. B. Giannakis are with the Dept. of ECE and the Digital Technology Center, University of Minnesota, 117 Pleasant Str. SE, Minneapolis, MN 55455. Te:(612) 625-4287; Emails: \texttt{\{shenx513,baing011,georgios\}@umn.edu }}

\vspace{-8mm}
\begin{abstract}
Structural equation models (SEMs) and vector autoregressive models (VARMs) are two broad families of approaches that have been shown useful in effective brain connectivity studies. While VARMs postulate that a given region of interest in the brain is directionally connected to another one by virtue of time-lagged influences, SEMs assert that causal dependencies arise due to contemporaneous effects, and may even be adopted when nodal measurements are not necessarily multivariate time series. To unify these complementary perspectives, linear structural vector autoregressive models (SVARMs) that leverage both contemporaneous and time-lagged nodal data have recently been put forth. Albeit simple and tractable, linear SVARMs are quite limited since they are incapable of modeling nonlinear dependencies between neuronal time series. To this end, the overarching goal of the present paper is to considerably broaden the span of linear SVARMs by capturing nonlinearities through kernels, which have recently emerged as a powerful nonlinear modeling framework in canonical machine learning tasks, e.g., regression, classification, and dimensionality reduction. The merits of kernel-based methods are extended here to the task of learning the effective brain connectivity, and an efficient regularized estimator is put forth to leverage the edge sparsity inherent to real-world complex networks. Judicious kernel choice from a preselected dictionary of kernels is also addressed using a data-driven approach. Extensive numerical tests on ECoG data captured through a study on epileptic seizures demonstrate that it is possible to unveil previously unknown causal links between brain regions of interest.
\end{abstract}

\begin{IEEEkeywords}
Network topology inference, structural vector autoregressive models, nonlinear models	
\end{IEEEkeywords}


\section{Introduction}
\label{S:1}

Several contemporary studies in the neurosciences have converged on the well-accepted view that information processing capabilities of the brain are facilitated by the existence of a complex underlying network; see e.g., \cite{rubinov2010complex} for a comprehensive review. The general hope is that understanding the behavior of the brain through the lens of network science will reveal important insights, with an enduring impact on applications in both clinical and cognitive neuroscience. 

However, brain networks are not directly observable, and must be inferred from observable or measurable neuronal processes. To this end, functional magnetic resonance imaging (fMRI) has emerged as a powerful tool, capable of revealing varying blood oxygenation patterns modulated by brain activity \cite{roche2009pioneering}. Other related brain imaging modalities include positron emission tomography (PET), electroencephalography (EEG), and electrocorticography (ECoG), to name just a few. Most state-of-the-art tools for inference of brain connectivity leverage variants of causal and correlational analysis methods, applied to time-series obtained from the imaging modalities \cite{friston1993time,buchel1997modulation,goebel2003investigating,gitelman2003modeling,deshpande2008effective}.

Contemporary brain connectivity analyses fall under two broad categories, namely, \emph{functional connectivity} which pertains to discovery of non-directional pairwise correlations between regions of interest (ROIs), and \emph{effective connectivity} which instead focuses on inference of directional (a.k.a., causal) dependencies between them~\cite{friston1994functional}. Granger causality or vector autoregressive models (VARMs)~\cite{roebroeck2005mapping}, structural equation models (SEMs)~\cite{mclntosh1994structural}, and dynamic causal modeling (DCM)~\cite{friston2003dynamic} constitute widely used approaches for effective connectivity studies. VARMs postulate that connected ROIs exert time-lagged dependencies among one another, while SEMs assume instantaneous causal interactions among them. Interestingly, these points of view are unified through the so-termed structural vector autoregressive model (SVARM)~\cite{chen2011vector}, which postulates that the spatio-temporal behavior observed in brain imaging data results from both instantaneous and time-lagged interactions between ROIs. It has been shown that SVARMs lead to markedly more flexibility and explanatory power than VARMs and SEMs treated separately, at the expense of increased model complexity.


The fundamental appeal of the aforementioned effective connectivity approaches stems from their inherent simplicity, since they adopt linear models. However, this is an oversimplification that is highly motivated by the need for tractability, even though consideration of nonlinear models for causal dependence may lead to more accurate approaches for inference of brain connectivity. In fact, recognizing the limitations associated with linear models, several variants of nonlinear SEMs have been put forth in a number of recent works; see e.g.,~\cite{joreskog1996nonlinear,wall2000estimation,jiang2010bayesian,lee2003model,harring2012comparison,kelava2014nonlinear}. 

For example,~\cite{lee2003model} and~\cite{lee2003maximum} advocate SEMs in which nonlinear dependencies only exist in the so-termed \emph{exogenous} or independent variables. Furthermore,~\cite{jiang2010bayesian} puts forth a hierarchical Bayesian nonlinear modeling approach in which unknown random parameters capture the strength and directions of causal links among variables. Several other studies adopt polynomial SEMs, which offer an immediate extension to classical linear SEMs; see e.g., \cite{joreskog1996nonlinear,wall2000estimation,harring2012comparison,kelava2014nonlinear}. In all these contemporary approaches, it is assumed that the network connectivity structure is \emph{known a priori}, and developed algorithms only estimate the unknown edge weights. However, this is a rather major limitation since such prior information may not be available in practice, especially when dealing with potentially massive networks, e.g., the brain. 

Similarly, several variants of nonlinear VARMs have been shown useful in unveiling links that often remain undiscovered by traditional linear models; see e.g.,\cite{marinazzo2008kernel,marinazzo2008kernel2,sun2008assessing,lim2015operator}. More recently, \cite{lim2015operator} proposed a kernel-based VARM, with nonlinear dependencies among nodes encoded by unknown functions belonging to a reproducing kernel Hilbert space.

Building upon these prior works, the present paper puts forth a novel additive nonlinear VARM to capture dependencies between observed ROI-based time-series, without explicit knowledge of the edge structure. Similar to \cite{lim2015operator}, kernels are advocated as an encompassing framework for nonlinear learning tasks. Note that SVARMs admit an interesting interpretation as SEMs, with instantaneous terms viewed as \emph{endogenous} variables, and time-lagged terms as exogenous variables. Since numerical measurement of external brain stimuli is often impractical, or extremely challenging in conventional experiments, adoption of such a fully-fledged SEM (with both endo- and exogenous inputs) is often impossible with traditional imaging modalities. 

A key feature of the novel approach is the premise that edges in the unknown network are sparse, that is, each ROI is linked to only a small subset of all potential ROIs that would constitute a maximally-connected power graph. This sparse edge connectivity has recently motivated the development of efficient regularized estimators, promoting the inference of sparse network adjacency matrices; see e.g.,~\cite{BGG14,angelosante2011sparse,kopsinis2011online,lim2015operator} and references therein. Based on these prior works, this paper develops a sparse-regularized kernel-based nonlinear SVARM to estimate the effective brain connectivity from per-ROI time series. Compared with~\cite{lim2015operator}, the novel approach incorporates instantaneous (cf. endogenous for SEMs) variables, it turns out to be more computationally efficient than~\cite{lim2015operator}, and facilitates a data-driven approach for kernel selection.

The rest of this paper is organized as follows. Section~\ref{sec:method} introduces the conventional SVARM, while Section~\ref{sec:nmps} puts forth its novel nonlinear variant. Section~\ref{sec:kernels} advocates a sparsity-promoting regularized least-squares estimator for topology inference from the nonlinear SVARM, while Section \ref{sec:mkl} deals with an approach to learn the kernel that `best' matches the data. Results of extensive numerical tests based on EEG data from an Epilepsy
study are presented in Section~\ref{sec:test}, and pertinent comparisons with linear variants demonstrate the efficacy of the novel approach. Finally, Section~\ref{sec:conclusion} concludes the paper, and highlights several potential future research directions opened up by this work.

\noindent\textit{Notation}. Bold uppercase (lowercase) letters will denote matrices (column vectors), while operators $(\cdot)^{\top}$, and $\textrm{diag}(\cdot)$ will stand for matrix transposition and diagonal matrices, respectively. The identity matrix will be represented by $\mathbf{I}$, while $\mathbf{0}$ will denote the all-zero matrix, and their dimensions will be clear from the context. Finally, $\ell_p$ and Frobenius norms will be denoted by $\|\cdot\|_p$, and $\|\cdot\|_F$, respectively.

\begin{figure*}[tpb!]
\begin{minipage}[b]{.49\textwidth}
\centering
\includegraphics[width=7.3cm]{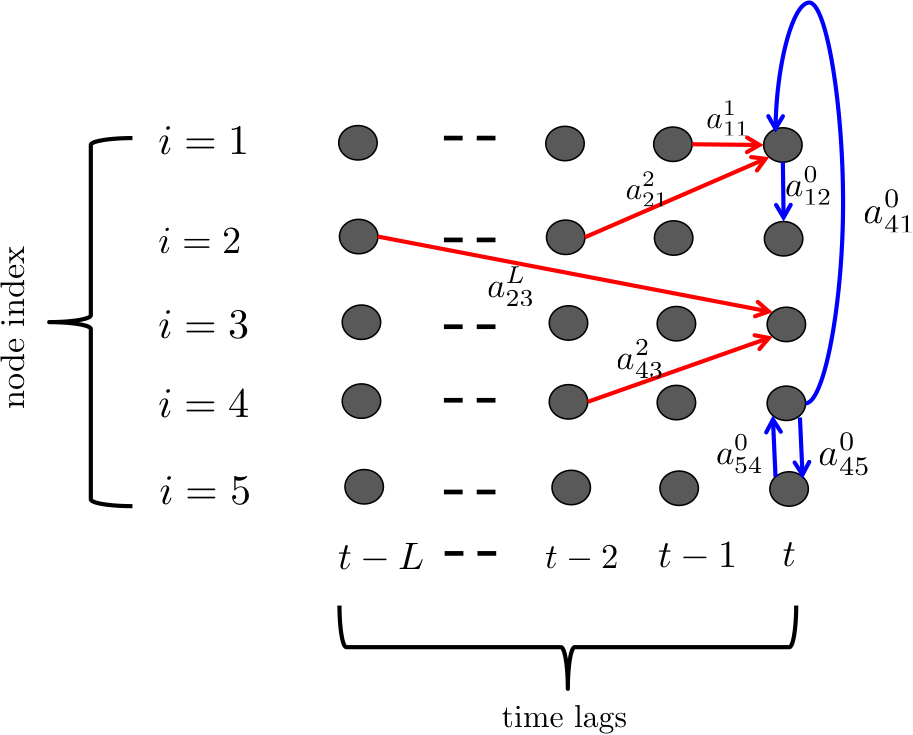}
\end{minipage}
\begin{minipage}[b]{.49\textwidth}
\centering
\includegraphics[width=8cm]{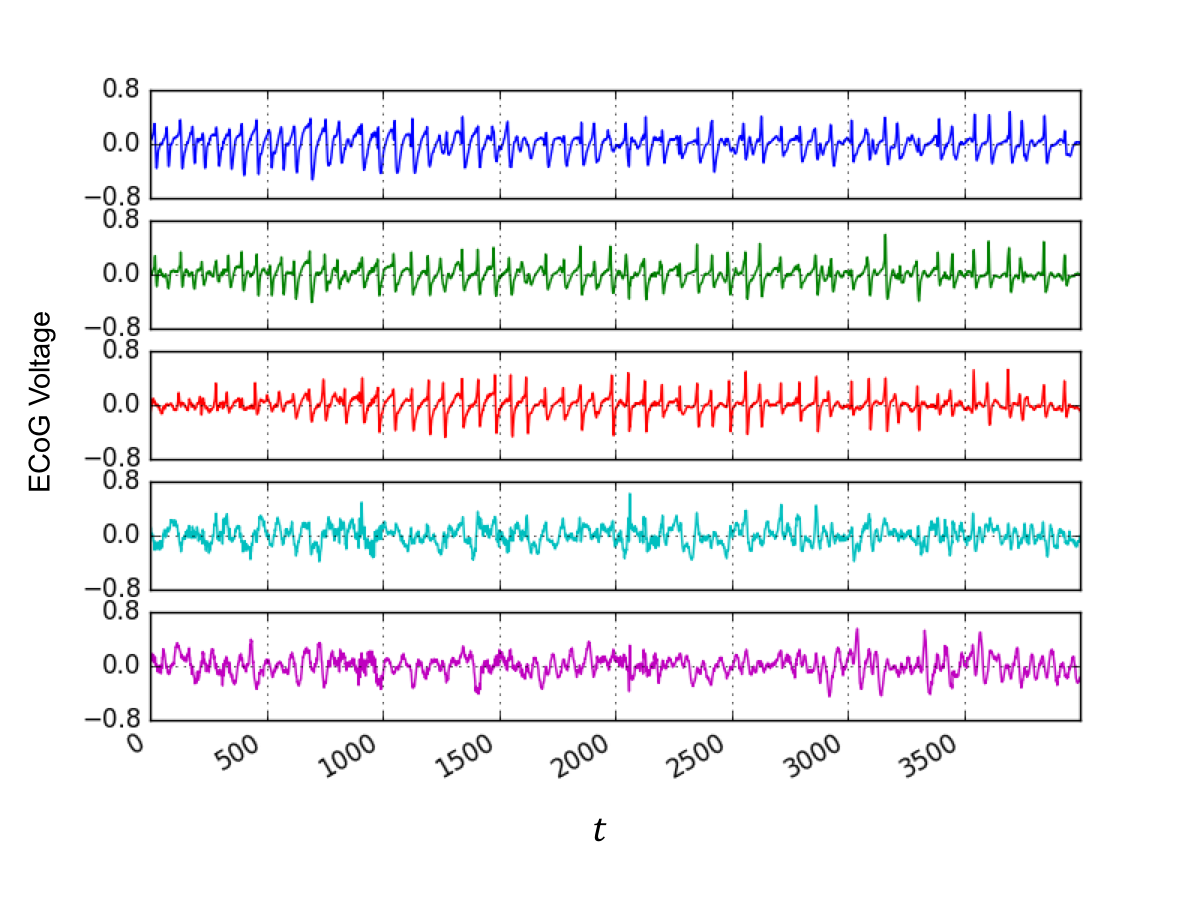}
\end{minipage}
 \caption{(left) A simple illustration of a $5$-node brain network; and (right) a set of five neuronal time series (e.g., ECoG voltage) each associated with a node. Per interval $t$, SVARMs postulate that causal dependencies between the $5$ nodal time series may be due to both the instantaneous effects (blue links), and/or time-lagged effects (red links). Estimating the values of the unknown coefficients amounts to learning the causal (link) structure of the network.} 
 \label{fig:svar}
\end{figure*}

\section{Preliminaries on Linear SVARMs}
\label{sec:method}

Consider a directed network whose topology is unknown, comprising $N$ nodes, each associated with an observable time series $\{ y_{it} \}_{t=1}^T$ measured over $T$ time-slots, for $i=1,\dots,N$. Note that $y_{it}$ denotes the $t$-th sample of the time series measured at node $i$. In the context of the brain, each node could represent a ROI, while the per-ROI time series are obtainable from standard imaging modalities, e.g., EEG or fMRI time courses. The network topology or edge structure will be captured by the weighted graph adjacency matrix $\mathbf{A} \in \mathbb{R}^{N \times N}$, whose $(i,j)$-th entry $a_{ij}$ is nonzero only if a directed (causal) effect exists from region $i$ to region $j$.

In order to unveil the hidden causal network topology, traditional linear SVARMs postulate that each $y_{jt}$ can be represented as a linear combination of instantaneous measurements at other nodes $\{ y_{it} \}_{i \neq j}$, and their time-lagged versions $ \{ \{ y_{i(t-\ell)}\}_{i=1}^N \}_{\ell = 1}^L$~ \cite{chen2011vector}. Specifically, $y_{jt}$ admits the following linear instantaneous plus time-lagged model
 \begin{align}
 	y_{jt}= \sum_{i\neq j}a_{ij}^{0}y_{it} + \sum_{i=1}^N\sum_{\ell=1}^L a_{ij}^{\ell} y_{j(t-\ell)} +e_{jt}
	\label{eq:svar:linear}
 \end{align}
with $a_{ij}^{\ell}$ capturing the causal influence of region $i$ upon region $j$ over a lag of $\ell$ time points, while $a_{ij}^{0}$ encodes the corresponding instantaneous causal relationship between them. The coefficients encode the causal structure of the network, that is, a causal link exists between nodes $i$ and $j$ only if $a_{ij}^{0} \neq 0$, or if there exists $a_{ij}^{\ell} \neq 0$ for $\ell = 1, \dots, L$. If $a_{ij}^{0} = 0 \; \forall i, j$, then~\eqref{eq:svar:linear} reduces to classical Granger causality \cite{roebroeck2005mapping}. Similarly, setting $a_{ij}^{\ell} = 0 \; \forall i, j, \ell\neq 0$ reduces~\eqref{eq:svar:linear} to a linear SEM with no exogenous inputs~\cite{kaplan09}.
Defining $\bby_t := \left[ y_{1t}, \dots, y_{Nt} \right]^{\top}$, $\bbe_t := \left[e_{1t}, \dots, e_{Nt}\right]^{\top}$, and the time-lagged adjacency matrix $\bbA^{\ell} \in \mathbb{R}^{N \times N}$ with the $(i,j)$-th entry $\left[ \bbA^{\ell} \right]_{ij} := a_{ij}^{\ell}$, one can write~\eqref{eq:svar:linear} in vector form as 
\begin{equation}
\label{eq:svar_vec}
\bby_t = \bbA^0 \bby_t + \sum\limits_{\ell=1}^L \bbA^{\ell} \bby_{t-\ell} + \bbe_t
\end{equation}
where $\bbA^{0}$ has zero diagonal entries $a_{ii}^{0} = 0$ for $i=1,\dots,N$.

Given the multivariate time series $\{ \bby_t \}_{t=1}^T$, the goal is to estimate matrices $\{ \bbA^{\ell} \}_{\ell=0}^L$, and consequently unveil the hidden network topology. Although generally known, one can readily deduce $L$ via standard model order selection tools e.g., the Bayesian information criterion \cite{chen1998speaker}, or Akaike's information criterion~\cite{bozdogan1987model}. 

Knowing which entries of $\bbA^0$ are nonzero, several approaches have been put forth to estimate their values. Examples are based upon ordinary least-squares~\cite{chen2011vector}, and hypothesis tests developed to detect presence or absence of pairwise causal links under prescribed false-alarm rates~\cite{roebroeck2005mapping}. Albeit conceptually simple and computationally tractable, the linear SVARM is incapable of capturing nonlinear dependencies inherent to complex networks such as the human brain. To this end, the present paper generalizes the \emph{linear} SVARM in~\eqref{eq:svar:linear} to a \emph{nonlinear} kernel-based SVARM.

It is also worth noting that most real world networks (including the brain) exhibit edge sparsity, the tendency for each node to link with only a few other nodes compared to the maximal $\mathcal{O}(N)$ set of potential connections per node. This means that per $j$, only a few coefficients $\{ a_{ij}^{\ell}\}$ are nonzero. In fact, several recent approaches exploiting edge sparsity have been advocated, leading to more efficient topology estimation; see e.g., ~\cite{BGG14,angelosante2011sparse,lim2015operator}. 

\section{From linear to nonlinear SVARMs}
\label{sec:nmps}

To enhance flexibility and accuracy, this section generalizes~\eqref{eq:svar:linear} so that nonlinear causal dependencies can be captured. The most general nonlinear model with both the instantaneous and time-lagged structure can be written in multivariate form as $\bby_{t}=\bar{\bbf}(\bby_{-jt}, \{\bby_{t-\ell}\}_{\ell=1}^L)+\bbe_t$, or, entry-wise as
\begin{align}
\label{eq:svar:general}
	y_{jt}=\bar{f}_j(\bby_{-jt}, \{\bby_{t-\ell}\}_{\ell=1}^L)+e_{jt},\quad j=1,\ldots, N
\end{align}
where $\bby_{-jt}:=[y_{1t}, \ldots, y_{(j-1)t}, y_{(j+1)t}, \ldots, y_{Nt}]^\top$ collects all but the $j$-th nodal observation at time $t$, $\bby_{t-\ell}:=[y_{1(t-\ell)}, \ldots, y_{N(t-\ell)}]^\top$, and $\bar{f}_j(.)$ denotes a nonlinear function of its multivariate argument. With limited $(NT)$ data available, $\bar{f}_j$ in \eqref{eq:svar:general} entails $(L+1)N-1$ variables. This fact motivates simpler model functions to cope with the emerging `curse of dimensionality' in estimating $\{\bar{f}_j\}_{j=1}^N$. A simplified form of~\eqref{eq:svar:general} has been studied in~\cite{lim2015operator} with $L=1$, and without instantaneous influences $\bby_{-jt}$, which have been shown of importance in applications such as brain connectivity~\cite{mclntosh1994structural} and gene regulatory networks~\cite{cai2013inference}. Such a model is simplified compared with \eqref{eq:svar:general} because the number of variables of $\bar{f}_j$ reduces to $N$. Nevertheless, estimating such an $N$-variate functional model still suffers from the curse of dimensionality, especially when the size of typical networks scales up.

To circumvent this challenge, we further posit that the multivariate function in \eqref{eq:svar:general} is separable with respect to each of its $(L+1)N-1$ variables.
Such a simplification of \eqref{eq:svar:general} amounts to adopting a generalized additive model (GAM)~\cite[Ch. 9]{trevor2011elements}. In the present context, the GAM adopted is $\bar{f}_j(\bby_{-jt}, \{\bby_{t-\ell}\}_{\ell=1}^L)=\sum_{i\neq j} \bar{f}^0_{ij}(y_{it}) + \sum_{i=1}^N\sum_{\ell=1}^L  \bar{f}^{\ell}_{ij}  (y_{i(t-\ell)})$, where the nonlinear functions $\{\bar{f}_{ij}^\ell\}$ will be specified in the next section. Defining $\bar{f}_{ij}^{\ell}(y):=a_{ij}^\ell f_{ij}^\ell (y)$, the node $j$ observation at time $t$ is a result of both instantaneous and multi-lag effects; that is [cf. \eqref{eq:svar:linear}]
\begin{align}
\label{eq:svar:nonlinear}
	y_{jt}=\sum_{i\neq j}a^0_{ij} f^0_{ij}(y_{it}) &+ \sum_{i=1}^N\sum_{\ell=1}^L a^{\ell}_{ij}   f^{\ell}_{ij}  (y_{i(t-\ell)}) +e_{jt}
\end{align}
where similar to \eqref{eq:svar:linear}, $\{a_{ij}^\ell\}$ define the matrices $\{\bbA^\ell\}_{\ell=0}^L$. As before, a directed edge from node $j$ to node $i$ exists if the corresponding $a_{ij}^{\ell}\neq 0$ for any $\ell =0, 1, \ldots, L$. Instead of having to estimate an $[(L+1)N-1]$-variate function in \eqref{eq:svar:general} or an $N$-variate function in \cite{lim2015operator}, \eqref{eq:svar:nonlinear} requires estimating $(L+1)N-1$ \emph{univariate} functions. Note that conventional linear SVARMs in \eqref{eq:svar:linear} assume that the functions $\{f_{ij}^{\ell}\}$ in \eqref{eq:svar:nonlinear} are linear, a limitation that the ensuing Section~\ref{sec:kernels} will address by resorting to a reproducing kernel Hilbert space (RKHS) formulation to model $\{f_{ij}^{\ell}\}$.  

\noindent\textbf{Problem statement.} Given $\{\bby_{t}\in\mathbb{R}^N\}_{t=1}^T$, the goal now becomes to estimate the nonlinear functions $\{f_{ij}^{\ell}\}$, as well as the adjacency matrices $\{\bbA^\ell\}_{\ell=0}^L$ in \eqref{eq:svar:nonlinear}.

\section{Kernel-based Sparse SVARMs}
\label{sec:kernels}
Suppose that each univariate function $f_{ij}^{\ell}(.)$ in \eqref{eq:svar:nonlinear} belongs to the RKHS
\begin{align}
\label{def:rkhs}
	\mathcal{H}_i^{\ell}:=\{f_{ij}^{\ell}|f_{ij}^{\ell}(y)=\sum_{t=1}^{\infty} \beta_{ijt}^{\ell}\kappa_i^{\ell}(y, y_{i(t-\ell)})\}	
\end{align}
where $\kappa_i^{\ell}(y,\psi): \mathbb{R}\times \mathbb{R}\rightarrow \mathbb{R}$ is a preselected basis (so-termed kernel) function that measures the similarity between $y$ and $\psi$. Different choices of $\kappa_i^\ell$ specify their own basis expansion spaces, and the linear functions can be regarded as a special case associated with the linear kernel  $\kappa_i^\ell(y,\psi)=y\psi$.  An alternative popular kernel is the Gaussian one that is given by $\kappa_i^\ell(y, \psi):= \exp[-(y-\psi)^2/(2\sigma^2)]$. A kernel is reproducing if it satisfies $\langle \kappa_{i}^\ell(y, \psi_1), \kappa_i^\ell(y,\psi_2)\rangle=\kappa_i^{\ell}(\psi_1,\psi_2)$, which induces the RKHS norm $\|f_{ij}^\ell\|_{\mathcal{H}_{i}^\ell}^2=\sum_{\tau}\sum_{\tau'} \beta^\ell_{ij\tau}\beta^\ell_{ij\tau'}\kappa_i^{\ell}(y_{i\tau},y_{i\tau'})$.

Considering the measurements per node $j$, with functions $f_{ij}^{\ell}\in \mathcal{H}_i^l$, for $i=1,\ldots, N$ and $\ell=0,1,\ldots, L$, the present paper advocates the following regularized least-squares (LS) estimates of the aforementioned functions obtained as
\begin{align}
\label{eq:prob:0}
	&\{\hat{f}_{ij}^{\ell}\}=\arg \min_{\{f_{ij}^{\ell}\in \mathcal{H}_i^{\ell}\}}  \frac{1}{2}\sum_{t=1}^T\bigg[y_{jt}-\sum_{i\neq j}a^0_{ij} f^0_{ij}(y_{it})\nonumber\\
	&- \sum_{i=1}^N\sum_{\ell=1}^L a^{\ell}_{ij}   f^{\ell}_{ij}  (y_{it}) \bigg]^2+\lambda \sum_{i=1}^N\sum_{\ell=0}^L\Omega(\|a_{ij}^\ell f_{ij}^\ell\|_{\mathcal{H}_i^\ell})
\end{align}
where $\Omega(.)$ denotes a regularizing function, which will be specified later. An important result that will be used in the following is the representer theorem~\cite[p.\;169]{trevor2011elements}, according to which the optimal solution for each $f_{ij}^\ell$ in \eqref{eq:prob:0} is given by
\begin{align}
\label{eq:sol:fij}
	\hat{f}_{ij}^\ell(y)=&\sum_{t=1}^T \beta_{ijt}^{\ell}\kappa_i^{\ell}(y,y_{i(t-\ell)}).
\end{align}
Although the function spaces in~\eqref{def:rkhs} include infinite basis expansions, since the given data are finite, namely $T$ per node, the optimal solution in \eqref{eq:sol:fij} entails a finite basis expansion.
Substituting \eqref{eq:sol:fij} into \eqref{eq:prob:0}, and letting $\bbbeta_{ij}^\ell:=[\beta_{ij1}^\ell, \ldots, \beta_{ijT}^\ell]^\top$, and $\bbalpha_{ij}^\ell:=a_{ij}^\ell \bbbeta_{ij}^\ell$, the functional minimization in \eqref{eq:prob:0} boils down to optimizing over vectors $\{\bbalpha_{ij}^\ell\}$. Specifically, \eqref{eq:prob:0} can be equivalently written in vector form as
\begin{align}
\label{eq:obj:1}
	&\{\hat{\bbalpha}_{ij}^\ell\}=\arg \min_{\{\bbalpha_{ij}^{\ell}\}} \frac{1}{2}\bigg\|\bby_{j}-\sum_{i\neq j}\bbK_{i}^{0}\bbalpha_{ij}^0 
	- \sum_{i=1}^N\sum_{\ell=1}^L \bbK_{i}^{\ell}\bbalpha_{ij}^\ell  \bigg\|_2^2\nonumber\\
	&\hspace{3cm}+\lambda \sum_{i=1}^N\sum_{\ell=0}^L\Omega\bigg(\sqrt{(\bbalpha_{ij}^{\ell})^\top\bbK_i^{\ell}\bbalpha_{ij}^{\ell}}\bigg)
\end{align}
where $\bby_j:=[y_{j1}, \ldots, y_{jT}]^\top$, and the $T\times T$ matrices $\{\bbK_i^\ell\}$ have entries $[\bbK_i^\ell]_{t,\tau}=\kappa_i^\ell(y_{it}, y_{i(\tau-\ell)})$.
Furthermore, collecting all the observations at different nodes in $\bbY:=[\bby_1, \ldots, \bby_N] \in \mathbb{R}^{T\times N}$ and letting $\bar{\bbK}^{\ell}:=[\bbK^{\ell}_1 \dots \bbK^{\ell}_N]$, \eqref{eq:obj:1} can be written as
\begin{multline}
\label{eq:obj:ls}
	\{\bbalpha_{ij}^\ell\}=\arg \min_{\bbalpha_{ii}^0=\mathbf{0},\{\bbalpha_{ij}^{\ell}\}} \frac{1}{2} \bigg \|\bbY-\sum_{l=1}^L\bar{\bbK}^{\ell} \bbW_{\alpha}^{\ell} \bigg \|_F^2\\
	+\lambda \sum_{i=1}^N\sum_{\ell=0}^L\Omega\bigg(\sqrt{(\bbalpha_{ij}^{\ell})^\top\bbK_i^{\ell}\bbalpha_{ij}^{\ell}}\bigg) 	
\end{multline} 
where the $NT \times N$ block matrix 
\begin{align}
\bbW^{\ell}_{\alpha}:=\left[
\begin{array}{cll}
\bbalpha_{11}^{\ell} &	\cdots &\bbalpha_{1N}^{\ell}\\
\vdots & \ddots & \vdots\\
\bbalpha_{N1}^{\ell} &\cdots	& \bbalpha_{NN}^{\ell}\end{array} \right]\label{def:W}
\end{align}
exhibits a structure `modulated' by the entries of $\mathbf{A}^{\ell}$. For instance, if $a_{ij}^{\ell}=0$, then $\bbalpha_{ij}^{\ell}:=a_{ij}^\ell \bbbeta_{ij}^\ell$ is an all-zero block, irrespective of the values taken by $\bbbeta_{ij}^{\ell}$. 

Instead of the LS cost used in \eqref{eq:prob:0} and \eqref{eq:obj:ls}, alternative loss functions could be employed to promote robustness using the $\epsilon$-insensitive, or, the $\ell_1$-error norm; see e.g., \cite[Ch.\;12]{trevor2011elements}. Regarding the regularizing function $\Omega(.)$, typical choices are $\Omega(z)=|z|$, or, $\Omega(z)=z^2$.  The former is known to promote sparsity of edges, which is prevalent to most networks; see e.g., \cite{rubinov2010complex}. In principle, leveraging such prior knowledge naturally leads to more efficient topology estimators, since $\{\bbA^{\ell}\}$ are promoted to have only a few nonzero entries. The sparse nature of $\bbA^{\ell}$ manifests itself as \emph{block sparsity} in $\bbW^{\ell}_{\alpha}$. Specifically, using $\Omega(z)=|z|$, one obtains the following estimator of the coefficient vectors $\{\bbalpha_{ij}^\ell\}$ for $j=1, \ldots, N$
\begin{align}
\label{eq:prob:mod}
\{\hat{\bbalpha}_{ij}^{\ell} \} =& \underset{ \hat{\bbalpha}_{ii}^{0}=\boldsymbol{0},\{\bbalpha_{ij}^{\ell}\}}{\text{arg min}}  \;\; 
	\frac{1}{2} \bigg \|\bbY-\sum_{l=1}^L\bar{\bbK}^{\ell} \bbW_{\alpha}^{\ell} \bigg \|_F^2\nonumber \\
&\hspace{1cm}+\lambda\sum_{\ell=0}^L \sum_{j=1}^N\sum_{i=1}^N\sqrt{ (\bbalpha_{ij}^{\ell})^\top\bbK_i^{\ell}\bbalpha_{ij}^{\ell}}.
\end{align} 
Recognizing that summands in the regularization term of~\eqref{eq:prob:mod} can be written as $\sqrt{ (\bbalpha_{ij}^{\ell})^\top\bbK^{\ell}_i\bbalpha^{\ell}_{ij}} = \| (\bbK^{\ell}_i)^{1/2} \bbalpha^{\ell}_{ij} \|_2$, which is the weighted $\ell_2$-norm of $\bbalpha_{i,j}$, the entire regularizer can henceforth be regarded as the weighted $\ell_{2,1}$-norm of $\bbW_{\alpha}^\ell$, that is known to be useful for promoting block sparsity. It is clear that~\eqref{eq:prob:mod} is a convex problem, which admits a globally optimal solution. In fact, the problem structure of~\eqref{eq:prob:mod} lends itself naturally to efficient iterative \emph{proximal optimization} methods e.g., proximal gradient descent iterations~\cite[Ch. 7]{bertsekas}, or, the alternating direction method of multipliers (ADMM) \cite{schizas3}. 

For a more detailed description of algorithmic approaches adopted to unveil the hidden topology by solving~\eqref{eq:prob:mod}, the reader is referred to Appendix~\ref{sec:algs}. All in all,  Algorithm \ref{alg:admm} is a summary of the novel iterative solver of \eqref{eq:prob:mod} derived based on ADMM iterations. Per iteration, the complexity of ADMM is in the order of $\mathcal{O}(T^2NL)$, which is linear in the network size $N$. A couple of remarks are now in order.
\begin{remark}
Selecting $\Omega(z)=z^2$ is known to control model complexity, and thus prevent overfitting \cite[Ch. 3]{trevor2011elements}. Let $\bbD^{\ell}{:=}\text{Bdiag}(\bbK_1^{\ell} \dots \bbK_N^{\ell})$, and $\bbD{:=}\text{Bdiag}(\bbD^{0} \dots \bbD^{L})$, where 
$\text{Bdiag}(.)$ is a block diagonal of its matrix arguments. Substituting  $\Omega(z)=z^2$ into \eqref{eq:obj:ls}, one obtains
\begin{multline}
\label{eq:obj:ridge}
	\{\hat{\bbalpha}_{ij}^{\ell} \} = \underset{ \hat{\bbalpha}_{ii}^{0}=\boldsymbol{0},\{\bbalpha_{ij}^{\ell}\}}{\text{arg min}}  \;\;\frac{1}{2} \bigg \|\bbY-\bar{\bbK}\bbW_{\alpha} \bigg \|_F^2 \\
	+\lambda ~\text{trace}(\bbW_{\alpha}^\top\bbD\bbW_{\alpha})
\end{multline}
where $\bar{\bbK}:=[\bar{\bbK}^{0} \ldots\bar{\bbK}^{L}]$, and $\bbW_{\alpha}:=[(\bbW_{\alpha}^{0})^\top \ldots (\bbW_{\alpha}^{L})^\top]^\top$. Problem \eqref{eq:obj:ridge} is convex and can be solved in closed form as
\begin{align}
\label{sol:ridge}
	\hat{\bar{\bbalpha}}_j=\left(\bar{\bbK}_j^\top\bar{\bbK}_j+2\bbD_j\right)^{-1}\bar{\bbK}_j^\top \bby_j
\end{align}
where $\bar{\bbalpha}_j$ denotes the $(NL-1)T\times 1$ vector obtained after removing entries of the $j$-th column of $\bbW_{\alpha}$ indexed by $\mathcal{I}_j:=\{(j-1)T+1, \ldots, jT\}$; $\bar{\bbK}_j$ collects columns of $\bar{\bbK}$ excluding the columns indexed by $\mathcal{I}_j$; and the block-diagonal matrix $\bbD_j$ is obtained after eliminating rows and columns of $\bbD$ indexed by $\mathcal{I}_j$.	 Using the matrix inversion lemma, the complexity of solving \eqref{sol:ridge} is in the order of $\mathcal{O}(T^3NL)$.
\end{remark}

\begin{remark}
\label{remark2}
Relying on an operator kernel (OK), 
the approach in \cite{lim2015operator} offers a more general nonlinear VARM (but not SVARM) than the one adopted here. However,  \cite{lim2015operator} did not account for instantaneous or the multiple-lagged effects. Meanwhile, estimating $\bar{\bbf}(\bby_{t-1})$ in~\cite{lim2015operator} does not scale well as the size of the network ($N$) increases. Also OK-VARM is approximated in \cite{lim2015operator} using the Jacobian, which again adds to the complexity of the algorithm, and may degrade the generality of the proposed model. Finally, the model in \cite{lim2015operator} is limited in its ability to incorporate the structure of the network (e.g., edge sparsity). In order to incorporate prior information on the model structure, \cite{lim2015operator} ends up solving a nonconvex problem, which might experience local minima, and the flexibility in choosing kernel functions will also be sacrificed. In contrast, our approach entails a natural extension to a data-driven kernel selection, which will be outlined in the next section.
\end{remark}
\begin{algorithm}[t] 
	\caption{ADMM for network topology identification}\label{alg:admm}
	\begin{algorithmic} [1]
		\State\textbf{Input:}~$\bbY$, $\{ \{ \bbK_i^{\ell} \}_{i=1}^N \}_{\ell=1}^L$, $\tau_{\alpha}$, $\lambda$, $\rho$
		
		\State\textbf{Initialize:}~$ \boldsymbol{\Gamma}[0]=\boldsymbol{0}_{NT\times N}$, $ \boldsymbol{\Xi}[0]=\boldsymbol{0}_{NT\times N}$, $k=0$
		
		\For{$ \ell = 1, \ldots, L$}
			\State $\bbD^{\ell}{=}\text{Bdiag}(\bbK_1^{\ell}, \dots, \bbK_N^{\ell})$  
			\State $\bar{\bbK}^{\ell}=[\bbK_1^{\ell} \ldots \bbK_N^{\ell}]$ 
		\EndFor
		
		\State $\bar{\bbK}:=[\bar{\bbK}^{0}  \ldots \bar{\bbK}^{L}]$, $\bbD{=}\text{Bdiag}(\bbD^{0} \dots \bbD^L)$
		
		\For{$j = 1, \ldots, N$}
			\State $\mathcal{I}_j:=\{(j-1)T+1, \ldots, jT\}$
	        \State $\bar{\mathcal{I}}_j :=\{(k,l)|k\notin \mathcal{I}_j, \text{or } l\notin\mathcal{I}_j \} $ 
	        \State $\tilde{\mathcal{I}}_j :=\{(k,l)| l\notin\mathcal{I}_j \}$
	        \State $ \bbD_j = \left[ \bbD \right]_{\bar{\mathcal{I}}_j}$, $\bar{\bbK}_j = \big[ \bar{\bbK} \big]_{\tilde{\mathcal{I}}_j}$
		\EndFor

		\While {\text{not converged}}
			\For{$j=1,\ldots , N$ (in parallel)}
			\State $\bbq_j[k]=\rho \bbD^{1/2}_j\bbgamma_j[k]+\bar{\bbK}_j^\top\bby_j-\bbD_j^{1/2}\bbxi_{j}[k]$
		\State $\bar{\bbalpha}_{j}[k+1]=\left(\bar{\bbK}_j^\top \bar{\bbK}_j+\rho \bbD_j\right)^{-1}\bbq_j[k]$
		
		\State $\bbgamma_{ij}^{\ell}[k]=\mathcal{P}_{\lambda/\rho}\left((\bbK_i^{\ell})^{1/2}\bbalpha_{ij}^{\ell}[k+1]+\bbxi_{ij}^{\ell}[k]/\rho\right)$,
		
		\State \hspace{1mm}for $i=1,\ldots N$, $\ell=0, \ldots L$

		\EndFor		
		
		\State $\bbW_{\alpha}[k+1]:=[(\bbW_{\alpha}^{0})^\top[k+1], \ldots, (\bbW_{\alpha}^{L})^\top[k+1]]^\top$
		\State $\bbGamma[k+1]:=[(\bbGamma^{0}[k+1])^\top, \ldots, (\bbGamma^{L}[k+1])^\top]^\top$ 
		\State $ \boldsymbol{\Xi}[k+1]  =  \boldsymbol{\Xi}[k]+\rho (\bbD^{1/2}\bbW_{\alpha}[k+1]-\bbGamma[k+1])$
		\State $k = k + 1$
		\EndWhile
		\State \textbf{Edge identification:}(after converging to $\hat{\bbalpha}^*_{ij}$)\\
		$~\hat{a}^*_{ij}\neq 0$ if $\|\hat{\bbalpha}^*_{ij}\|\geq \tau_\alpha$, else $\hat{a}^*_{ij}=0$, $\forall ~(i, j)$ \\
		\Return $\{\hat{\bbA^\ell}\}_{\ell=0}^L$ 		
	\end{algorithmic}
\end{algorithm}

\section{Data-driven kernel selection}
\label{sec:mkl}
Choice of the kernel function determines the associated Hilbert space, and it is therefore of significant importance in estimating the nonlinear functions $\{f_{ij}^\ell\}$. Although Section \ref{sec:kernels}  assumed that the kernels $\{\kappa_i^\ell\}$ are available, this is not the case in general, and this section advocates a data-driven strategy for selecting them. Given a dictionary of reproducing kernels $\{\kappa_p\}_{p=1}^P$, it has been shown that any function in the convex hull $\mathcal{K}:=\{\kappa| \kappa=\sum_{p=1}^P\theta_p \kappa_p,~ \theta_p\geq 0, ~\sum_{p=1}^P \theta_p=1\}$ is a reproducing kernel. Therefore, the goal of the present section is to select a kernel from $\mathcal{K}$ that best fits the data. For ease of exposition, consider $\kappa_i^\ell=\kappa \in\mathcal{K} $, for all $\ell=0, 1,\ldots, L$ and $i=1, \ldots, N$ in \eqref{eq:prob:0}, therefore $\mathcal{H}_i^\ell=\mathcal{H}^{(\kappa)}$. Note that the formulation can be readily extended to settings when $\{\kappa_i^\ell\}$ are different. Incorporating  $\kappa$ as a variable function in \eqref{eq:prob:0} yields 
\begin{align}
	\label{eq:prob:mkl0}
	\{\hat{f}_{ij}^{\ell}\}&=\arg \min_{\kappa\in \mathcal{K},\{f_{ij}^{\ell}\in \mathcal{H}^{(\kappa)}\}} \frac{1}{2}\sum_{t=1}^T \bigg[y_{jt}-\sum_{i\neq j}a^0_{ij} f^0_{ij}(y_{it})\nonumber\\
	&\hspace{-10mm}- \sum_{i=1}^N\sum_{\ell=1}^L a^{\ell}_{ij}   f^{\ell}_{ij}  (y_{it}) \bigg]^2+\lambda \sum_{i=1}^N\sum_{\ell=0}^L\Omega(\|a_{ij}^\ell f_{ij}^\ell\|_{\mathcal{H}^{(\kappa)}})
\end{align}
where $\mathcal{H}^{(\kappa)}$ denotes the Hilbert space associated with kernel function $\kappa$. With $\mathcal{H}_p$ denoting the RKHS induced by $\kappa_p$, it has been shown in \cite{bazerque2013nonparametric} and \cite{micchelli2005learning} that the optimal $\{\hat{f}_{ij}^\ell\}$ in \eqref{eq:prob:mkl0} is expressible in a separable form as
\begin{align}
\label{eq:mkl:fij}
\hat{f}_{ij}^\ell(y): =\sum_{p=1}^Pf_{ij}^{\ell,p}(y)
\end{align}
where $f_{ij}^{\ell,p}$ belongs to RKHS $\mathcal{H}_p$, for $p=1, \ldots, P$. Substituting \eqref{eq:mkl:fij} into \eqref{eq:prob:mkl0}, one obtains
\begin{align}
	\label{eq:prob:mkl1}
	&\{\hat{f}_{ij}^{\ell}\}=\arg   \min_{\{f_{ij}^{\ell,p}\in \mathcal{H}_p\}}
	\frac{1}{2}\sum_{t=1}^T \bigg[y_{jt}-\sum_{i\neq j}\sum_{p=1}^P a^0_{ij}f^{0,p}_{ij}(y_{it})\\
	&- \sum_{i=1}^N\sum_{\ell=1}^L\sum_{p=1}^P a^{\ell}_{ij}   f^{\ell,p}_{ij}  (y_{it}) \bigg]^2
	+\lambda \sum_{i=1}^N\sum_{\ell=0}^L\sum_{p=1}^P \Omega(\|a_{ij}^\ell f_{ij}^{\ell,p}\|_{\mathcal{H}_p}).\nonumber
\end{align}
Note that \eqref{eq:prob:mkl1} and \eqref{eq:prob:0} have similar structure, and their only difference pertains to an extra summation over $P$ candidate kernels. Hence, \eqref{eq:prob:mkl1} can be solved in an efficient manner along the lines of the iterative solver of \eqref{eq:prob:0} listed under Algorithm \ref{alg:admm} [cf. the discussion in Section~\ref{sec:kernels}]. Further details of the solution are omitted due to space limitations.

\section{Numerical tests}
\label{sec:test}
This section presents test results on seizure data, captured through experiments conducted in an \emph{epilepsy} study~\cite{kramer2008emergent}. Epilepsy refers to a chronic neurological condition characterized by recurrent seizures, globally afflicting over $20$ million people, and often associated with abnormal neuronal activity within the brain. Diagnosis of the condition sometimes involves comparing EEG or ECoG time series obtained from a patient's brain before and after onset of a seizure. Recent studies have shown increasing interest in analysis of connectivity networks inferred from the neuronal time series, in order to gain more insights about the unknown physiological mechanisms underlying epileptic seizures. In this section, connectivity networks are inferred from the seizure data using the novel approach, and a number of comparative measures are computed from the identified network topologies.

\subsection{Seizure data description}
\label{subsec:dataset}
Seizure data were obtained for a $39$-year-old female subject with a case of intractable epilepsy at the \emph{University of California, San Francisco (UCSF) Epilepsy Center}; see also~\cite{kramer2008emergent}. An $8 \times 8$ subdural electrode grid was implanted into the cortical surface of the subject's brain, and two accompanying electrode strips, each comprising six electrodes (a.k.a., depth electrodes) were implanted deeper into the brain. Over a period of five days, the combined electrode network recorded $76$ ECoG time series, consisting of voltage levels measured in a region within close proximity of each electrode. 

ECoG epochs containing eight seizures were extracted from the record and analyzed by a specialist. The time series at each electrode were first passed through a bandpass filter, with cut-off frequencies of $1$ and $50$ Hz, and the so-termed \emph{ictal} onset of each seizure was identified as follows. A board-certified neurophysiologist identified the initial manifestation of rhythmic high-frequency, low-voltage focal activity, which characterizes the onset of a seizure. Samples of data before and after this seizure onset were then extracted from the ECoG time series. The per-electrode time series were then divided into $1$s windows, with $0.5$s overlaps between consecutive windows, and the average spectral power between $5$Hz and $15$Hz was computed per window. Finally, power spectra over all electrodes were averaged, and the ictal onset was identified by visual inspection of a dramatic increase in the average power. Two temporal intervals of interest were picked for further analysis, namely, the \emph{preictal} and ictal intervals. The preictal interval is defined as a $10$s interval preceding seizure onset, while the ictal interval comprises the $10$s immediately afterwards. Further details about data acquisition and pre-processing are provided in~\cite{kramer2008emergent}.
  
The goal here was to assess whether modeling nonlinearities, and adopting the novel kernel-based approach would yield significant insights pertaining to causal/effective dependencies between brain regions, that linear variants would otherwise fail to capture. Toward this goal, several standard network analysis measures were adopted to characterize the structural properties of the inferred networks. 

\begin{figure}[tbp!]
\begin{minipage}[t]{.24\textwidth}
\centering
\includegraphics[width=4.5cm]{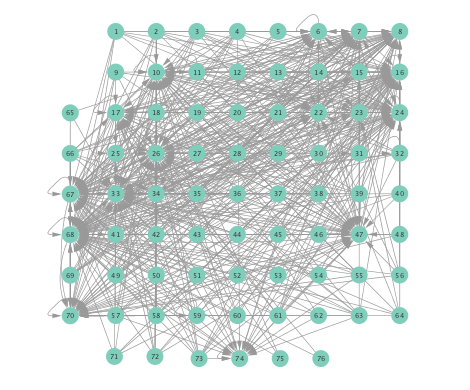}
\centerline{(a)}
\end{minipage}
\begin{minipage}[t]{.24\textwidth}
\centering
\includegraphics[width=4.5cm]{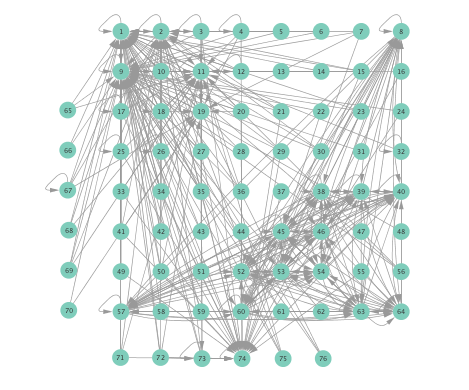}
\centerline{(b)}
\end{minipage}
\begin{minipage}[t]{.24\textwidth}
\centering
\includegraphics[width=4.5cm]{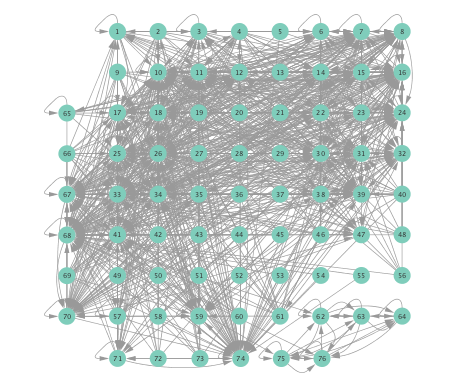}
\centerline{(c)}
\end{minipage}
\begin{minipage}[t]{.24\textwidth}
\centering
\includegraphics[width=4.5cm]{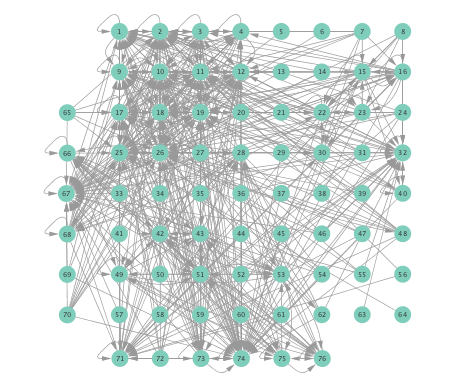}
\centerline{(d)}
\end{minipage}
\begin{minipage}[t]{.24\textwidth}
\centering
\includegraphics[width=4.5cm]{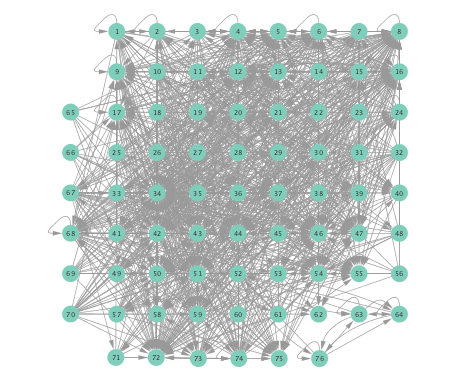}
\centerline{(e)}
\end{minipage}
\begin{minipage}[t]{.24\textwidth}
\centering
\includegraphics[width=4.5cm]{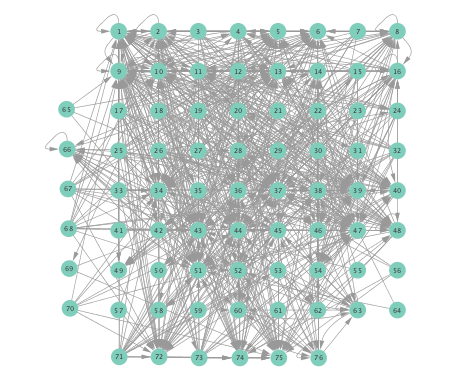}
\centerline{(f)}
\end{minipage}
 \caption{Visualizations of $76$-electrode networks inferred from ECoG data: (a) linear SVARM with $L=1$ on preictal time series; (b) linear SVARM with $L=1$ on ictal time series; (c) K-SVARM with $L=1$ on preictal time series, using a polynomial kernel of order $2$;  (d) the same K-SVARM on ictal time series; (e) K-SVARM with kernel selection on preictal time series; and finally (f) K-SVARM with kernel selection on ictal time series.} 
 \label{fig:top}
\end{figure}

\subsection{Inferred networks}
\label{subsec:inferred}
Prior to running the developed algorithm, $10$s intervals were chosen from the preprocessed ECoG data, and then divided into $20$ successive segments, each spanning $0.5$s. To illustrate this, suppose the $10$s interval starts from $t=0$s and ends at $t=10s$, then the first segment comprises samples taken over the interval $[0s, 0.5s]$, the second one would be $[0.5s, 1s]$, and so on. After this segmentation of the time series, directed network topologies were inferred using Algorithm~\ref{alg:admm} with $L=1$, based on the $0.5$s segments, instead of the entire signal, to ensure that the signal is approximately \emph{stationary} per experiment run. A directed link from electrode $i$ to $j$ was drawn if at least one of the estimates of $a_{ij}^\ell$ turned out to be nonzero.

\begin{figure*}[t]
\begin{minipage}[t]{.24\textwidth}
\centering
\includegraphics[width=4.5cm]{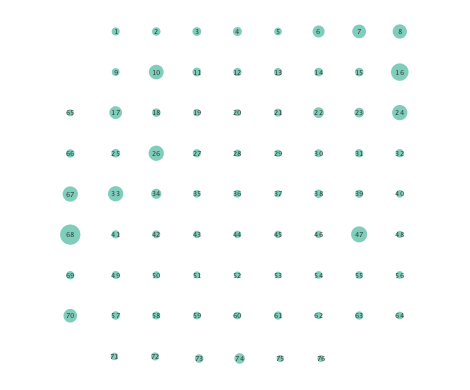}
\centerline{(a)}
\end{minipage}
\begin{minipage}[t]{.24\textwidth}
\centering
\includegraphics[width=4.5cm]{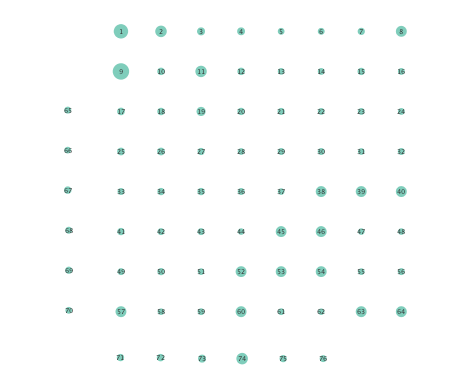}
\centerline{(b)}
\end{minipage}
\begin{minipage}[t]{.24\textwidth}
\centering
\includegraphics[width=4.5cm]{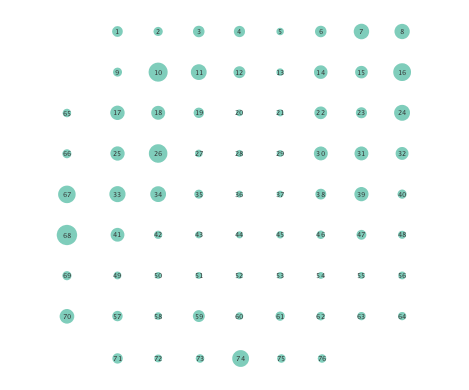}
\centerline{(c)}
\end{minipage}
\begin{minipage}[t]{.24\textwidth}
\centering
\includegraphics[width=4.5cm]{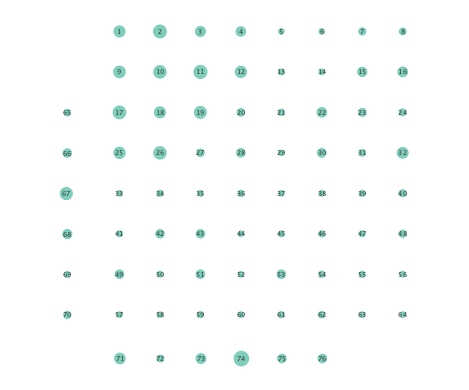}
\centerline{(d)}
\end{minipage}
 \caption{Node degrees of networks inferred from ECoG data encoded by circle radii: (a) linear SVARM on preictal data; (b)  linear SVARM on ictal data; (c) K-SVARM on preictal time series; and (d) K-SVARM on ictal data.} 
 \label{fig:dg}
\end{figure*}
Networks inferred from the preictal and ictal intervals were compared using the linear, the kernel-based (K-)SVARMs, as well as the K-SVARM with data-driven kernel selection. The lag lengths were set to $L=1$ for all cases. For the K-SVARM, a polynomial kernel of order $2$ was selected. Furthermore, the threshold in Algorithm~\ref{alg:admm} was set to $\tau_{\alpha} = 0.01$, $\rho$ was set to $0.01$, while the regularization parameter was selected via cross-validation. For the data-driven kernel selection scheme, two candidate kernels were employed, namely, a linear kernel, and a polynomial kernel of order $2$.

Figure \ref{fig:top} depicts networks inferred from different algorithms for both preictal and ictal intervals of the time series. The figure illustrates results obtained by the linear SVARM, and the K-SVARM approach with and without kernel selection. Each node in the network is representative of an electrode, and it is depicted as a circle, while the node arrangement is forced to remain consistent across the four visual representations. A cursory inspection of the visual maps reveals significant variations in connectivity patterns between ictal and preictal intervals for both models. Specifically, networks inferred via the K-SVARMs, reveal a global decrease in the number of links emanating from each node, while those inferred via the linear model depict increases and decreases in links connected to different nodes. Interestingly, the K-SVARM with kernel selection recovered most of the edges inferred by the linear and the K-SVARM using a polynomial kernel, which implies that both linear and nonlinear interactions may exist in brain networks. Clearly, one is unlikely to gain much insight only by visual inspection of the network topologies. To further analyze differences between inferred networks from both models, and to assess the potential benefits gained by adopting the novel scheme, several network topology metrics are computed and compared in the next subsection.

\subsection{Comparison of network metrics}
\label{subsec:net_measures}

First, in- and out-degree was computed for nodes in each of the inferred networks. Note that the in-degree of a node counts its number of incoming edges, while the out-degree counts the number of out-going edges. The total degree per node sums the in- and out-degrees, and is indicative of how well-connected a given node is. Figure~\ref{fig:dg} depicts nodes in the network and their total degrees encoded by the radii of circles associated with the nodes. As expected from the previous subsection, Figures~\ref{fig:dg} (a) and (b) demonstrate that the linear SVARM yields both increases and deceases in the inferred node degree. On the other hand, the nonlinear SVARM leads to a more spatially consistent observation with most nodes exhibiting a smaller degree after the onset of a seizure (see Figures~\ref{fig:dg} (c) and (d)), which may imply that causal dependencies thin out between regions of the brain once a seizure starts.

\begin{figure*}[t]
\begin{minipage}[t]{.24\textwidth}
\centering
\includegraphics[width=4.2cm]{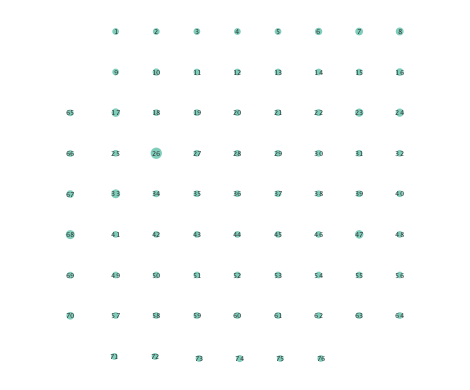}
\end{minipage}
\begin{minipage}[t]{.24\textwidth}
\centering
\includegraphics[width=4.2cm]{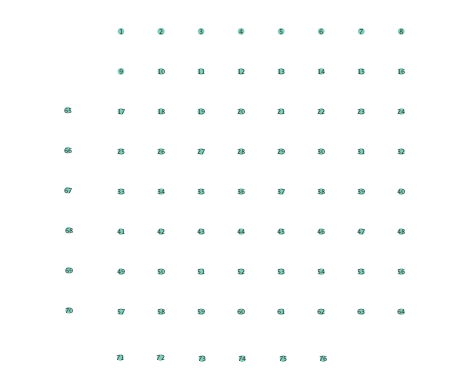}
\end{minipage}
\begin{minipage}[t]{.24\textwidth}
\centering
\includegraphics[width=4.2cm]{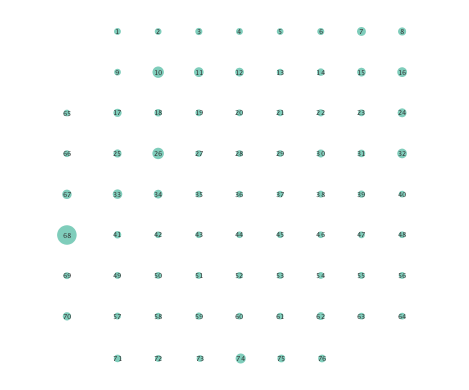}
\end{minipage}
\begin{minipage}[t]{.24\textwidth}
\centering
\includegraphics[width=4.2cm]{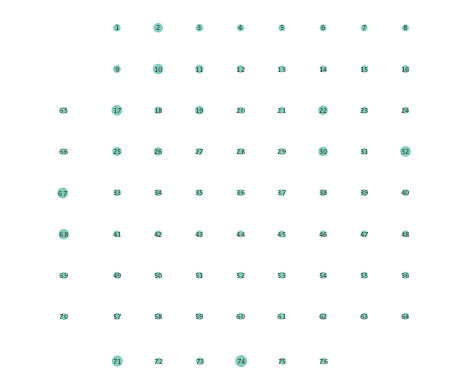}
\end{minipage}
 \caption{Same as in Figure \ref{fig:dg} for comparison based on betweenness centralities of inferred graphs.} 
 \label{fig:bc}
\end{figure*}

In order to assess the information-routing abilities of brain regions before and after seizure onset, comparisons of the so-termed \emph{betweenness centrality} were done. Betweenness centrality of a node computes the fraction of shortest paths between all node pairs that traverse the given node, and it is useful to identify the key information transmitting hubs in a network; see e.g.,~\cite{kolaczyk2009statistical} for more details. The per-node betweenness centrality for each inferred network are depicted in Figure \ref{fig:bc}, with node radii similarly encoding the computed values. Little variation between preictal and ictal betweenness centralities is seen for the linear model (Figures \ref{fig:bc} (a) and (b)), while variations are slightly more marked for the K-SVARM, see Figures~\ref{fig:bc} (c) and (d). It can be seen that modeling nonlinearities reveals subtle changes in information-routing capabilities of nodes between preictal and ictal phases.

\begin{figure*}[t]
\begin{minipage}[t]{.24\textwidth}
\centering
\includegraphics[width=4.2cm]{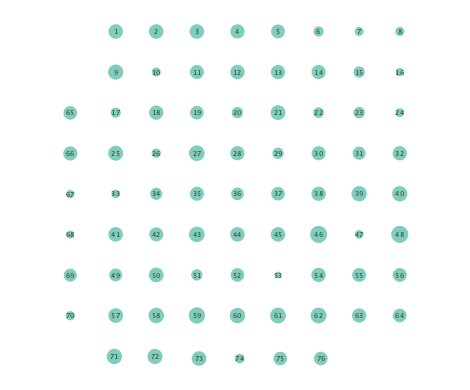}
\end{minipage}
\begin{minipage}[t]{.24\textwidth}
\centering
\includegraphics[width=4.2cm]{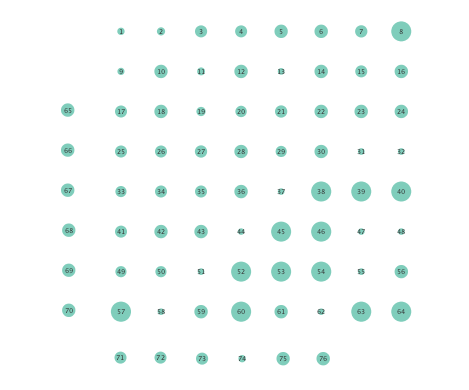}
\end{minipage}
\begin{minipage}[t]{.24\textwidth}
\centering
\includegraphics[width=4.2cm]{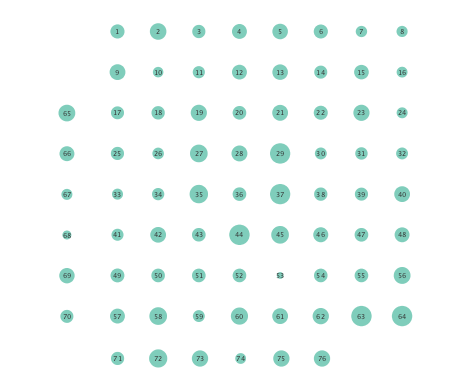}
\end{minipage}
\begin{minipage}[t]{.24\textwidth}
\centering
\includegraphics[width=4.2cm]{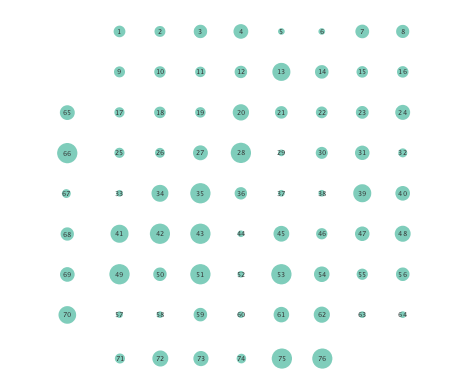}
\end{minipage}
 \caption{Same as in Figure \ref{fig:dg} for comparison based on clustering coefficients of inferred graphs.} 
 \label{fig:cc}
\end{figure*}
\begin{figure*}[t]
\begin{minipage}[t]{.24\textwidth}
\centering
\includegraphics[width=4.2cm]{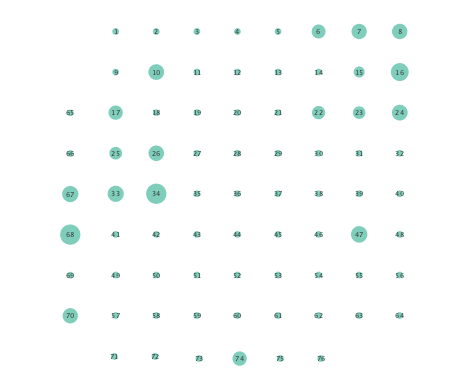}
\end{minipage}
\begin{minipage}[t]{.24\textwidth}
\centering
\includegraphics[width=4.2cm]{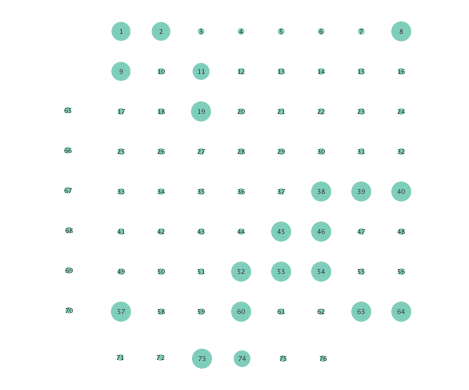}
\end{minipage}
\begin{minipage}[t]{.24\textwidth}
\centering
\includegraphics[width=4.2cm]{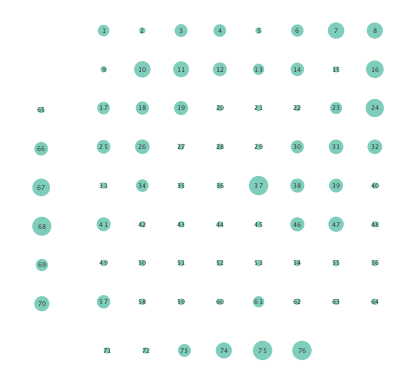}
\end{minipage}
\begin{minipage}[t]{.24\textwidth}
\centering
\includegraphics[width=4.2cm]{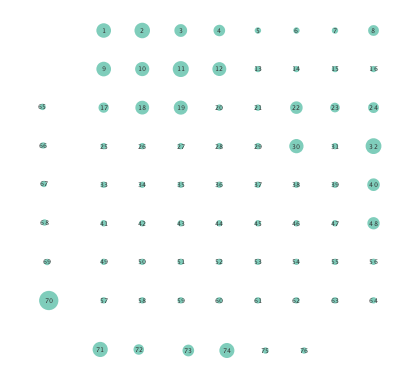}
\end{minipage}
 \caption{Same as in Figure \ref{fig:dg} for comparison based on closeness centrality of  inferred graphs.} 
 \label{fig:cloc}
\end{figure*}

\emph{Clustering coefficients} are generally used to quantify network cohesion, the tendency for nodes to form groups or communities. Comparison of such coefficients between the preictal and ictal phases may reveal differences in cohesive behavior after onset of a seizure. In the present paper, a per-node clustering coefficient is adopted, and it computes the fraction of triangles in which a node participates out of all possible triangles to which it could possibly belong~\cite{kolaczyk2009statistical}. Note that a triangle is defined as a fully connected three-node subgraph. Figure \ref{fig:cc} depicts clustering coefficients per electrode obtained during the ictal and preictal phases of the ECoG time series. While both the linear and nonlinear models yield changes in the computed coefficients, most nodes have lower clustering coefficients upon seizure onset in the networks inferred via the K-SVARM. 

Finally, Figure~\ref{fig:cloc} depicts the \emph{closeness centrality} computed per node in the inferred networks. Closeness centrality measures how reachable a node is from all other nodes, and is generally defined as the reciprocal of the sum of geodesic distances of the node from all other nodes in the network; see also \cite{kolaczyk2009statistical}. Once again, Figure~\ref{fig:cloc} depicts a more general decrease in closeness centralities after seizure onset in networks inferred by the nonlinear SVARM, as compared to the linear variant. This empirical result indicates a change in reachability between regions of the brain during an epileptic seizure.

\begin{figure*}[t]
\begin{minipage}[t]{.24\textwidth}
\centering
\includegraphics[width=4.2cm]{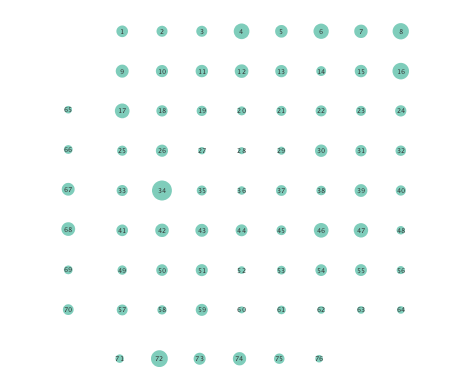}
\centerline{(a)}
\end{minipage}
\begin{minipage}[t]{.24\textwidth}
\centering
\includegraphics[width=4.2cm]{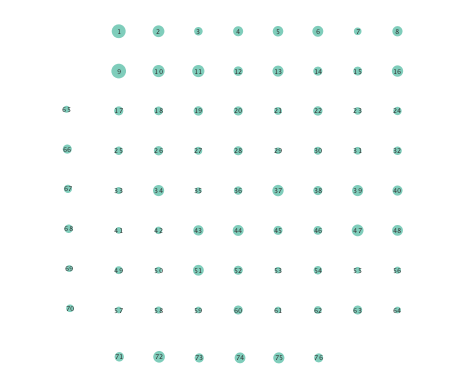}
\centerline{(b)}
\end{minipage}
\begin{minipage}[t]{.24\textwidth}
\centering
\includegraphics[width=4.2cm]{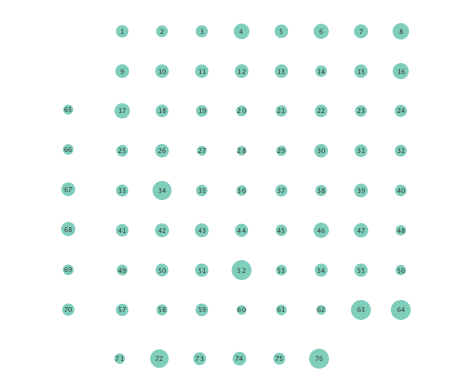}
\centerline{(c)}
\end{minipage}
\begin{minipage}[t]{.24\textwidth}
\centering
\includegraphics[width=4.2cm]{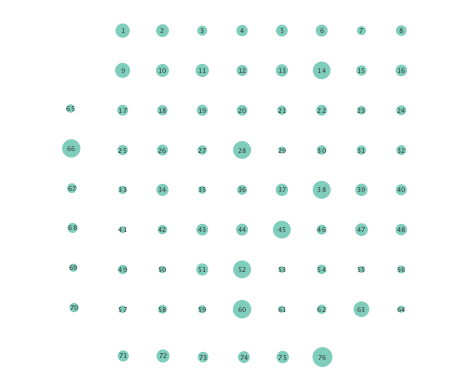}
\centerline{(d)}
\end{minipage}
 \caption{Comparison of node degrees of  inferred graphs: (a) K-SVARM with kernel selection on preictal data; (b)  K-SVARM with kernel selection on ictal data. Comparison of closeness centrality of  inferred graphs: (c) K-SVARM with kernel selection on preictal data; (d)  K-SVARM with kernel selection on ictal data.} 
 \label{fig:mkl}
\end{figure*}

Moreover, the performance of K-SVARM with data-driven kernel selection was also tested. Figure ~\ref{fig:mkl} illustrates the per node degree as well as the closeness centrality of networks inferred from preictal and ictal phases. Consistent with Figures \ref{fig:dg} and \ref{fig:cloc}, Figure \ref{fig:mkl} again reveals universal decrease in node degrees as well as closeness centrality at seizure onset.

In addition to the local metrics, a number of global measures were computed over entire inferred networks, and pertinent comparisons were drawn between the two phases; see Table~\ref{tab:net_metric} for a summary of the global measures of the inferred networks. Several global metrics were considered, e.g., network density, global clustering coefficient, network diameter, average number of neighbors, number of self loops, number of isolated nodes and the  size of the largest connected component. 

Network density refers to the number of actual edges divided by the number of potential edges, while the global clustering coefficient is the fraction of connected triplets that form triangles, adjusted by a factor of three to compensate for double counting. On the other hand, network diameter is the length of the longest geodesic, excluding infinity. Table \ref{tab:net_metric} shows that networks inferred via the K-SVARM exhibit lower network cohesion after seizure onset, as captured by network density, global clustering coefficient, and average number of neighbors, while the network diameter increases. 

These changes provide empirical evidence that the brain network becomes less connected, and diffusion of information is inhibited after the onset of an epileptic seizure. Also interestingly, it turns out that the number of self-loops significantly decrease during the ictal phase for networks inferred using the K-SVARM. Note that in this experiment, only connections to the previous time interval are considered ($L=1$), while $\bbA^0$ is constrained to have no self-loops. As a consequence, existence of a self loop reveals a strong temporal dependence between measurements at the same node. In fact, a drop in the number of self loops implies a lower temporal dependence between successive ECoG samples, a phenomenon that was not captured by the linear SVARM.

%
%

%
%

\begin{table}
\begin{center}
 \begin{tabular}[t]{ r | c | c| c| c }
\hline
\hline
 &  \multicolumn{2}{| c |}{\bf Linear SVARM} &  \multicolumn{2}{| c }{\bf K-SVARM}  \\ \hline
     & Preictal & Ictal  & Preictal  & Ictal  \\ \hline
   Network density  & 0.189  & 0.095 & 0.242 &0.148\\ \hline
   Glob. clustering coeff. & 0.716 & 0.731& 0.775 & 0.624\\ \hline
    No. of connect. comp. & 1 & 1 & 1 & 2\\ \hline
    Network diameter & 5  &  3 & 3&4\\ \hline    
    Avg. no. of neighbors&14.18 & 9.39 & 18.18&11.11\\ \hline
    No. of self loops & 19 & 30 & 42 &31 \\ \hline
    Size of Largest comp. & 76 & 76 &76 &67\\ \hline
\hline
    \end{tabular}
\end{center}
\caption{Comparison of global  metrics associated with networks inferred from ECoG seizure data using the linear and K-SVARMs with $L=1$. The K-SVARM was based on a polynomial kernel of order $2$. Major differences between the computed metrics indicate that one may gain insights from network topologies inferred via models that capture nonlinear dependencies.}\label{tab:net_metric}
\end{table}



\section{Conclusions}
\label{sec:conclusion}
This paper put forth a novel nonlinear SVARM framework that leverages kernels to infer effective connectivity networks in the brain. Postulating a generalized additive model with unknown functions to capture the hidden network structure, a novel regularized LS estimator that promotes sparse solutions was advocated. In order to solve the ensuing convex optimization problem, an efficient algorithm that resorts to ADMM iterations was developed, and a data-driven approach was introduced to select the appropriate kernel. Extensive numerical tests were conducted on ECoG seizure data from a study on epilepsy. 

In order to assess the utility of the novel approach, several local and global metrics were adopted and computed on networks inferred before and after the onset of a seizure. By observing changes in network behavior that are revealed by standard metrics before and after seizure onset, it is possible identify key structural differences that may be critical to explain the mysteries of epileptic seizures. With this in mind, the paper focused on identifying structural differences in the brain network that could not be captured by the simpler linear model. Interestingly, empirical results support adoption of a nonlinear modeling perspective when analyzing differences in effective brain connectivity for epilepsy patients. Specifically, adopting the novel kernel-based approach revealed more significant differences between the preictal and ictal phases of ECoG time series. For instance, it turned out that some regions exhibited fewer dependencies, reduced reachability, and weakened information-routing capabilities after the onset of a seizure. Since the kernel-based model includes the linear SVARM as an instance, the conducted experiments suggest that one may gain more insights by adopting the nonlinear model, a conclusion that may yield informative benefits to studies of epilepsy that leverage network science. 

This work paves the way for a number of exciting research directions in analysis of brain networks. Although it has been assumed that inferred networks are static, overwhelming evidence suggests that topologies of brain networks are dynamic, and may change over rather short time horizons. Future studies will extend this work to facilitate tracking of dynamic brain networks. Furthermore, the novel approach will be empirically tested on a wider range of neurological illnesses and disorders, and pertinent comparisons will be done to assess the merits of adopting the advocated nonlinear modeling approach. 

\section*{Appendix}
\renewcommand{\thesubsection}{\Alph{subsection}}

\subsection{Topology Inference via ADMM}
\label{sec:algs}

Given matrices $\bbY$ and $\bar{\bbK}:=[\bar{\bbK}^{0} \ldots\bar{\bbK}^{L}]$, this section capitalizes on convexity, and the nature of the additive terms in~\eqref{eq:prob:mod} to develop an efficient topology inference algorithm. Proximal optimization approaches have recently been shown useful for convex optimization when the cost function comprises the sum of smooth and nonsmooth terms; see e.g.,~\cite{daubechies2004iterative}. Prominent among these approaches is the alternating direction method of multipliers (ADMM), upon which the novel algorithm is based; see e.g., \cite{schizas3} for an early application of ADMM to distributed estimation.

For ease of exposition, let the equality constraints ($\bbalpha_{jj}^{\ell}=\boldsymbol{0}$) temporarily remain implicit. Introducing the change of variables $\bbgamma_{ij}^{\ell}: = (\bbK_{i}^{\ell})^{1/2}\bbalpha_{ij}^\ell $, problem~\eqref{eq:prob:mod} can be recast as
\begin{align}
\label{eq:prob:admm}
\nonumber
 \underset{\{\bbalpha_{ij}^{\ell}\} }{\text{arg min}} &
	(1/2) \|\bbY-\sum_{\ell=0}^L \bar{\bbK}^{\ell} \bbW_{\alpha}^{\ell}\|_F^2+\sum_{\ell=0}^Lg(\bbGamma^{\ell}) \\
\text{s.t.}~~ &  \bbgamma_{ij}^{\ell}-(\bbK_{i}^{\ell})^{1/2}\bbalpha_{ij}^{\ell}=\boldsymbol{0} ~\;\;\; \forall i, j,\ell
\end{align}
where $ \bbGamma^{\ell}:=[\bbgamma^{\ell}_{1} \ldots \bbgamma^{\ell}_{N}]$, $\bbgamma_{j}^{\ell}:=[(\bbgamma_{1j}^{\ell})^\top \ldots(\bbgamma_{Nj}^{\ell})^\top]^\top$, and
$ g(\bbGamma^{\ell}):=\lambda \sum_{i=1}^N\sum_{j=1}^N\|\bbgamma_{ij}^{\ell}\|_2$ 
 is the nonsmooth regularizer.
Let $\bbD^{\ell}{:=}\text{Bdiag}(\bbK_1^{\ell} \dots \bbK_N^{\ell})$, and $\bbD{:=}\text{Bdiag}(\bbD^{0} \dots \bbD^{\ell})$, where 
$\text{Bdiag}(.)$ is a block diagonal of its matrix arguments. One can then write the augmented Lagrangian 
of~\eqref{eq:prob:admm} as
\begin{multline}
\label{eq:prob:aug_lagrangian}
	\mathcal{L}_{\rho}(\bbW_{\alpha},  \bbGamma, \boldsymbol{\Xi}) 
	=
	(1/2)\|\bbY-\bar{\bbK} \bbW_{\alpha}\|_F^2 +g(\bbGamma) \\
+\langle \boldsymbol{\Xi}, \bbD^{1/2} \bbW_{\alpha}-\bbGamma \rangle 
+ (\rho/2)\|\bbGamma - \bbD^{1/2}\bbW_{\alpha}\|_F^2
\end{multline}
where $\bbW_{\alpha}:=[(\bbW_{\alpha}^{0})^\top \ldots (\bbW_{\alpha}^{L})^\top]^\top$, and $\bbGamma:=[(\bbGamma^{0})^\top \ldots (\bbGamma^{L})^\top]^\top$. Note that $\boldsymbol{\Xi}$ is a matrix of dual variables that collects Lagrange multipliers corresponding to the equality constraints in~\eqref{eq:prob:admm}, $\langle \mathbf{P}, \mathbf{Q} \rangle$ denotes the inner product between $\mathbf{P}$ and $\mathbf{Q}$, while $\rho > 0$ a prescribed penalty parameter. ADMM boils down to a sequence of alternating minimization iterations to minimize $\mathcal{L}_{\rho}(\bbW_{\alpha},  \bbGamma, \boldsymbol{\Xi})$ over the primal variables $\bbW_{\alpha}$, and $\bbGamma$, followed by a gradient ascent step over the dual variables  $\boldsymbol{\Xi}$; see also~\cite{baingana2013dynamic,schizas3}. Per iteration $k+1$, this entails the following provably-convergent steps, see e.g. \cite{schizas3}
\begin{subequations}
	\begin{align}
	\label{admm:1}
     \bbW_{\alpha}[k+1] & =
     \underset{ \bbW_{\alpha} }{\arg \min}  \;\; \mathcal{L}_{\rho}(\bbW_{\alpha}, \bbGamma[k], \boldsymbol{\Xi}[k])\\
	\label{admm:2}
	\bbGamma[k+1] & =
	 \underset{\bbGamma}{\arg \min} \;\; \mathcal{L}_{\rho}(\bbW_{\alpha}[k+1], \bbGamma, \boldsymbol{\Xi}[k])\\
	\label{admm:3}
	\boldsymbol{\Xi}[k+1] & =
	 \boldsymbol{\Xi}[k]+\rho( \bbD^{1/2}\bbW_{\alpha}[k+1] - \bbGamma[k+1]).
	\end{align}
\end{subequations}
%

Focusing on $\bbW_{\alpha}[k+1] $, note that \eqref{admm:1} decouples across columns of $\bbW_{\alpha}$, and admits closed-form, parallelizable solutions. Incorporating the structural constraint $\bbalpha_{jj}^0=\boldsymbol{0}$, one obtains the following decoupled subproblem per column $j$
\begin{multline}
	\label{admm:obj:alpha}
	\bar{\bbalpha}_j[k+1] \\=  \arg \min_{\bar{\bbalpha}_j} \;\;\; (1/2) \bar{\bbalpha}_j^\top\left(\bar{\bbK}_j^\top\bar{\bbK}_j+\rho\bbD_j\right)\bar{\bbalpha}_j 
	-\bar{\bbalpha}_j^\top\bbq_j[k]
\end{multline}
where $\bbq_j[k]$ is constructed by removal of entries indexed by $\mathcal{I}_j$ from $\rho \bbD^{1/2}\bbgamma_j[k]+\bar{\bbK}^\top\bby_j-\bbD^{1/2}\bbxi_{j}[k]$, with $\bbxi_j[k]$ denoting the $j$-th column of $\bbXi[k]$. Assuming $\left(\bar{\bbK}_j^\top \bar{\bbK}_j+\rho \bbD_j\right)$ is invertible, the per-column subproblem~\eqref{admm:obj:alpha} admits the following closed-form solution per $j$
\begin{align}
\label{admm:alpha1}
	\bar{\bbalpha}_{j}[k+1]=&\left(\bar{\bbK}_j^\top \bar{\bbK}_j+\rho \bbD_j\right)^{-1}\bbq_j[k].
\end{align}
On the other hand,~\eqref{admm:2} can be solved per component vector $\bbgamma_{ij}^{\ell}$, and a closed-form solution can be obtained via the so-termed \emph{block shrinkage} operator for each $i$ and $j$, namely,
\begin{align}
\label{admm:gmma}
	\bbgamma_{ij}^{\ell}[k]=\mathcal{P}_{\lambda/\rho}\left((\bbK_i^{\ell})^{1/2}\bbalpha_{ij}^{\ell}[k+1]+\bbxi_{ij}^{\ell}[k]/\rho\right)
\end{align}
where $ \mathcal{ P}_{\lambda}(\bbz):=(\bbz/ \|\bbz \|_2) \max (\|\bbz\|_2-\lambda,0)$.
Upon convergence, $\{a_{ij}^{\ell}\}$ can be determined by thresholding $\hat{\bbalpha}_{ij}^{\ell}$, and declaring an edge present from $i$ to $j$, if there exists any $\hat{\bbalpha}_{ij}^{\ell}\neq \mathbf{0}$, for $\ell=1, \dots, L$.





\bibliographystyle{IEEEtranS}
\bibliography{net,myabrv,mybib}

\begin{thebibliography}{10}
\providecommand{\url}[1]{#1}
\csname url@samestyle\endcsname
\providecommand{\newblock}{\relax}
\providecommand{\bibinfo}[2]{#2}
\providecommand{\BIBentrySTDinterwordspacing}{\spaceskip=0pt\relax}
\providecommand{\BIBentryALTinterwordstretchfactor}{4}
\providecommand{\BIBentryALTinterwordspacing}{\spaceskip=\fontdimen2\font plus
\BIBentryALTinterwordstretchfactor\fontdimen3\font minus
  \fontdimen4\font\relax}
\providecommand{\BIBforeignlanguage}[2]{{%
\expandafter\ifx\csname l@#1\endcsname\relax
\typeout{** WARNING: IEEEtranS.bst: No hyphenation pattern has been}%
\typeout{** loaded for the language `#1'. Using the pattern for}%
\typeout{** the default language instead.}%
\else
\language=\csname l@#1\endcsname
\fi
#2}}
\providecommand{\BIBdecl}{\relax}
\BIBdecl

\bibitem{angelosante2011sparse}
D.~Angelosante and G.~B. Giannakis, ``Sparse graphical modeling of
  piecewise-stationary time series,'' in \emph{Proc. Int. Conf. Acoust. Speech
  Signal Process.}, Prague, Czech Republic, May 2011, pp. 1960--1963.

\bibitem{baingana2013dynamic}
B.~Baingana, G.~Mateos, and G.~B. Giannakis, ``Dynamic structural equation
  models for tracking topologies of social networks,'' in \emph{Proc. of
  Workshop on Computational Advances in Multi-Sensor Adaptive Processing},
  Saint Martin, Dec. 2013, pp. 292--295.

\bibitem{BGG14}
------, ``Proximal-gradient algorithms for tracking cascades over social
  networks,'' \emph{IEEE J. Sel. Topics Sig. Proc.}, vol.~8, no.~4, pp.
  563--575, Aug. 2014.

\bibitem{bazerque2013nonparametric}
J.~A. Bazerque and G.~B. Giannakis, ``Nonparametric basis pursuit via sparse
  kernel-based learning: A unifying view with advances in blind methods,''
  \emph{IEEE Signal Processing Magazine}, vol.~30, no.~4, pp. 112--125, Jul.
  2013.

\bibitem{bertsekas}
D.~P. Bertsekas, \emph{Nonlinear Programming}, 2nd~ed.\hskip 1em plus 0.5em
  minus 0.4em\relax Athena-Scientific, 1999.

\bibitem{bozdogan1987model}
H.~Bozdogan, ``Model selection and {A}kaike's information criterion ({AIC}):
  {T}he general theory and its analytical extensions,'' \emph{Psychometrika},
  vol.~52, no.~3, pp. 345--370, Sep. 1987.

\bibitem{buchel1997modulation}
C.~B{\"u}chel and K.~J. Friston, ``Modulation of connectivity in visual
  pathways by attention: {C}ortical interactions evaluated with structural
  equation modelling and f{MRI}.'' \emph{Cerebral Cortex}, vol.~7, no.~8, pp.
  768--778, Dec. 1997.

\bibitem{cai2013inference}
X.~Cai, J.~A. Bazerque, and G.~B. Giannakis, ``Inference of gene regulatory
  networks with sparse structural equation models exploiting genetic
  perturbations,'' \emph{PLoS Comp. Biol.}, vol.~9, no.~5, p. e1003068, May
  2013.

\bibitem{chen2011vector}
G.~Chen, D.~R. Glen, Z.~S. Saad, J.~P. Hamilton, M.~E. Thomason, I.~H. Gotlib,
  and R.~W. Cox, ``Vector autoregression, structural equation modeling, and
  their synthesis in neuroimaging data analysis,'' \emph{Computers in Biology
  and Medicine}, vol.~41, no.~12, pp. 1142--1155, Dec. 2011.

\bibitem{chen1998speaker}
S.~Chen and P.~Gopalakrishnan, ``Speaker, environment and channel change
  detection and clustering via the {B}ayesian information criterion,'' in
  \emph{Proc. DARPA Broadcast News Transcription and Understanding Workshop},
  Virginia, USA, Aug. 1998, pp. 127--132.

\bibitem{daubechies2004iterative}
I.~Daubechies, M.~Defrise, and C.~De~Mol, ``An iterative thresholding algorithm
  for linear inverse problems with a sparsity constraint,'' \emph{Comm. Pure
  Appl. Math.}, vol.~57, no.~11, pp. 1413--1457, Aug. 2004.

\bibitem{deshpande2008effective}
G.~Deshpande, X.~Hu, R.~Stilla, and K.~Sathian, ``Effective connectivity during
  haptic perception: A study using {G}ranger causality analysis of functional
  magnetic resonance imaging data,'' \emph{NeuroImage}, vol.~40, no.~4, pp.
  1807--1814, May 2008.

\bibitem{friston1993time}
K.~J. Friston, C.~D. Frith, and R.~S. Frackowiak, ``Time-dependent changes in
  effective connectivity measured with {PET},'' \emph{Human Brain Mapping},
  vol.~1, no.~1, pp. 69--79, Jan. 1993.

\bibitem{friston1994functional}
K.~J. Friston, ``Functional and effective connectivity in neuroimaging: A
  synthesis,'' \emph{Human Brain Map.}, vol.~2, no. 1-2, pp. 56--78, Jan. 1994.

\bibitem{friston2003dynamic}
K.~J. Friston, L.~Harrison, and W.~Penny, ``Dynamic causal modelling,''
  \emph{NeuroImage}, vol.~19, no.~4, pp. 1273--1302, Aug. 2003.

\bibitem{gitelman2003modeling}
D.~R. Gitelman, W.~D. Penny, J.~Ashburner, and K.~J. Friston, ``Modeling
  regional and psychophysiologic interactions in f{MRI}: The importance of
  hemodynamic deconvolution,'' \emph{NeuroImage}, vol.~19, no.~1, pp. 200--207,
  May 2003.

\bibitem{goebel2003investigating}
R.~Goebel, A.~Roebroeck, D.-S. Kim, and E.~Formisano, ``Investigating directed
  cortical interactions in time-resolved f{MRI} data using vector
  autoregressive modeling and {G}ranger causality mapping,'' \emph{Magnetic
  Resonance Imaging}, vol.~21, no.~10, pp. 1251--1261, Dec. 2003.

\bibitem{harring2012comparison}
J.~R. Harring, B.~A. Weiss, and J.-C. Hsu, ``A comparison of methods for
  estimating quadratic effects in nonlinear structural equation models.''
  \emph{Psychological Methods}, vol.~17, no.~2, pp. 193--214, Jun. 2012.

\bibitem{trevor2011elements}
T.~Hastie, R.~Tibshirani, and J.~Friedman, \emph{The {E}lements of
  {S}tatistical {L}earning}.\hskip 1em plus 0.5em minus 0.4em\relax Springer,
  2009.

\bibitem{jiang2010bayesian}
X.~Jiang, S.~Mahadevan, and A.~Urbina, ``Bayesian nonlinear structural equation
  modeling for hierarchical validation of dynamical systems,'' \emph{Mechanical
  Systems and Signal Processing}, vol.~24, no.~4, pp. 957--975, Apr. 2010.

\bibitem{joreskog1996nonlinear}
K.~G. J{\"o}reskog, F.~Yang, G.~Marcoulides, and R.~Schumacker, ``Nonlinear
  structural equation models: The {K}enny-{J}udd model with interaction
  effects,'' \emph{Advanced Structural Equation Modeling: Issues and
  Techniques}, pp. 57--88, Jan. 1996.

\bibitem{kaplan09}
D.~Kaplan, \emph{Structural Equation Modeling: Foundations and
  Extensions}.\hskip 1em plus 0.5em minus 0.4em\relax Sage, 2009.

\bibitem{kelava2014nonlinear}
A.~Kelava, B.~Nagengast, and H.~Brandt, ``A nonlinear structural equation
  mixture modeling approach for nonnormally distributed latent predictor
  variables,'' \emph{Structural Equation Modeling: A Multidisciplinary
  Journal}, vol.~21, no.~3, pp. 468--481, Jun. 2014.

\bibitem{kolaczyk2009statistical}
E.~D. Kolaczyk, \emph{Statistical Analysis of Network Data: Methods and
  Models}.\hskip 1em plus 0.5em minus 0.4em\relax Springer, 2009.

\bibitem{kopsinis2011online}
Y.~Kopsinis, K.~Slavakis, and S.~Theodoridis, ``Online sparse system
  identification and signal reconstruction using projections onto weighted
  $\ell_1$ balls,'' \emph{IEEE Trans. Sig. Proc.}, vol.~59, no.~3, pp.
  936--952, Mar. 2011.

\bibitem{kramer2008emergent}
M.~A. Kramer, E.~D. Kolaczyk, and H.~E. Kirsch, ``Emergent network topology at
  seizure onset in humans,'' \emph{Epilepsy Research}, vol.~79, no.~2, pp.
  173--186, May 2008.

\bibitem{lee2003model}
S.-Y. Lee and X.-Y. Song, ``Model comparison of nonlinear structural equation
  models with fixed covariates,'' \emph{Psychometrika}, vol.~68, no.~1, pp.
  27--47, Mar. 2003.

\bibitem{lee2003maximum}
S.-Y. Lee, X.-Y. Song, and J.~C. Lee, ``Maximum likelihood estimation of
  nonlinear structural equation models with ignorable missing data,'' \emph{J.
  of Educ. and Behavioral Stat.}, vol.~28, no.~2, pp. 111--134, Jun. 2003.

\bibitem{lim2015operator}
N.~Lim, F.~d'Alch{\'e} Buc, C.~Auliac, and G.~Michailidis, ``Operator-valued
  kernel-based vector autoregressive models for network inference,''
  \emph{Machine Learning}, vol.~99, no.~3, pp. 489--513, Jun. 2015.

\bibitem{marinazzo2008kernel}
D.~Marinazzo, M.~Pellicoro, and S.~Stramaglia, ``Kernel-{G}ranger causality and
  the analysis of dynamical networks,'' \emph{Physical Review E}, vol.~77,
  no.~5, p. 056215, May 2008.

\bibitem{marinazzo2008kernel2}
------, ``Kernel method for nonlinear {G}ranger causality,'' \emph{Physical
  Review Letters}, vol. 100, no.~14, p. 144103, Apr. 2008.

\bibitem{mclntosh1994structural}
A.~Mclntosh and F.~Gonzalez-Lima, ``Structural equation modeling and its
  application to network analysis in functional brain imaging,'' \emph{Human
  Brain Mapping}, vol.~2, no. 1-2, pp. 2--22, Oct. 1994.

\bibitem{micchelli2005learning}
C.~A. Micchelli and M.~Pontil, ``Learning the kernel function via
  regularization,'' \emph{Journal of Machine Learning Research}, vol.~6, no.
  Jul, pp. 1099--1125, Jul. 2005.

\bibitem{roche2009pioneering}
R.~Roche and S.~Commins, \emph{Pioneering {S}tudies in {C}ognitive
  {N}euroscience}.\hskip 1em plus 0.5em minus 0.4em\relax McGraw-Hill
  Education, 2009.

\bibitem{roebroeck2005mapping}
A.~Roebroeck, E.~Formisano, and R.~Goebel, ``Mapping directed influence over
  the brain using {G}ranger causality and f{MRI},'' \emph{NeuroImage}, vol.~25,
  no.~1, pp. 230--242, Mar. 2005.

\bibitem{rubinov2010complex}
M.~Rubinov and O.~Sporns, ``Complex network measures of brain connectivity:
  Uses and interpretations,'' \emph{NeuroImage}, vol.~52, no.~3, pp.
  1059--1069, Sep. 2010.

\bibitem{schizas3}
I.~D. Schizas, A.~Ribeiro, and G.~B. Giannakis, ``Consensus in ad hoc {WSNs}
  with noisy links {-Part I}: {D}istributed estimation of deterministic
  signals,'' \emph{IEEE Trans. Sig. Proc.}, vol.~56, pp. 350--364, Jan. 2008.

\bibitem{sun2008assessing}
X.~Sun, ``Assessing nonlinear {G}ranger causality from multivariate time
  series,'' in \emph{Proc. Eur. Conf. Mach. Learn. Knowl. Disc. Databases},
  Antwerp, Belgium, Sep. 2008, pp. 440--455.

\bibitem{wall2000estimation}
M.~Wall and Y.~Amemiya, ``Estimation for polynomial structural equation
  models,'' \emph{J. American Stat. Assoc.}, vol.~95, no. 451, pp. 929--940,
  Oct. 2000.

\end{thebibliography}







\end{document}